\documentclass[aps,prd]{revtex4}
\usepackage{graphicx}
\usepackage{amsmath}
\usepackage[all]{xy}
\usepackage{amssymb}
\newcommand{\be}{\begin{equation}}
\newcommand{\ee}{\end{equation}}
\newcommand{\ben}{\begin{eqnarray}}
\newcommand{\een}{\end{eqnarray}}
\newcommand{\bes}{\begin{subequations}}
\newcommand{\ees}{\end{subequations}}

\usepackage{graphicx,epsfig}
\usepackage{amsfonts}
\usepackage{amssymb}
\begin{document}
\title{Global defects in field theory with applications to condensed matter}

\author{D. Bazeia,$^1$ J. Menezes$^2$ and R. Menezes$^1$}
\affiliation{$^1$Departamento de F\'\i sica, Universidade Federal da Para\'\i ba\\
Caixa Postal 5008, 58051-970 Jo\~ao Pessoa, Para\'\i ba, Brazil\\
$^2$Centro de F\'\i sica do Porto, Departamento de F\'\i sica, Faculdade de Ci\^encias\\
Universidade do Porto, 4169-007 Porto, Portugal}

\date{\today}
\begin{abstract}
We review investigations on defects in systems described by real scalar
fields in $(D,1)$ space-time dimensions. We first work in one spatial dimension,
with models described by one and two real scalar fields, and in higher dimensions.
We show that when the potential assumes specific form, there are models which support stable global defects for $D$ arbitrary. We also show how to find first-order differential equations that solve the equations of motion, and how to solve models
in $D$ dimensions via soluble problems in $D=1$. We illustrate the procedure
examining specific models and showing how they may be used in applications
in different contexts in condensed matter physics, and in other areas.
\end{abstract}
\maketitle

\section{Introduction}

The search for defect structures of topological nature is of direct interest
to high energy physics\cite{ra,rs,vs} and condensed matter\cite{esche,wa}. In the present work, we shall be mainly interested in defect structures, as they appear in high energy physics. However, we shall investigate models which may be of good use in
applications in condensed matter, mainly in pattern formation and in magnetic systems, where one finds rich scenarios with defect structures playing important role to control some of their physical features.

In Field Theory, the search for defect structures started several years ago. In the 1960's, works of Hobard\cite{ho} and Derrick\cite{de} have shown that for standard real scalar field models, there is no way of generating stable defect structures unless one works in $(1,1)$ space-time dimensions. In one space dimension, the defect structure of topological nature is named kink. This result has led to the standard route to defect solutions, in which one enriches the model to support vortices, if one works with $(2,1)$ space-time dimensions, with the introduction of Abelian gauge field, coupled to charged scalar matter\cite{no}. Or to support monopoles, if one
works in $(3,1)$ space-time dimensions, with the introduction of non-Abelian gauge fields coupled to charged scalar matter\cite{th,po,jz}.

For models described by a single real scalar field, the Lagrange density has the form
\be
{\cal L}=\frac12\partial_\mu\phi\partial^\mu\phi-V(\phi)
\ee
We work in $(D,1)$ space-time dimensions, and the metric is $(+,-,\ldots,-)$, with
$x_\mu=(x_0=x^0=t,x_1=-x^1,x_2=-x^2,...,x_D=-x^D)$.  The equation of motion is
\be
\frac{\partial^2\phi}{\partial t^2}-\nabla^2\phi+\frac{dV}{d\phi}=0
\ee
For static field configuration, the equation of motion becomes
\be
\nabla^2\phi=\frac{dV}{d\phi}
\ee
The energy can be written as
\be
E=\int {dx^1dx^2...dx^D}\left[\frac12\nabla\phi\cdot\nabla\phi+V(\phi)\right]
\ee

In this work we shall be interested in models described by scalar fields, defined in $D=1$ and in $D\geq2.$ The case of a single spatial dimension is standard and will be investigated in the next Sect.~{\ref{sec:one}}. Extensions to two or more spatial dimensions will be considered in Sect.~{\ref{sec:two}}. 

In $D=1$ we consider models described by one field
\be\label{pot1}
V(\phi)=\frac12\, W^2_\phi
\ee
and by two fields\cite{bds,bnrt,bb00}
\be\label{pot2}
V(\phi,\chi)=\frac12\,\left(W^2_\phi+W^2_\chi\right)
\ee
Here $W$ is a smooth function of the fields, and $W_\phi=\partial W/\partial\phi,$ etc. For higher spatial dimensions, for $D\geq2$ we work with models described by the potentials\cite{bmm} 
\be\label{potD}
V(\phi;x)=\frac12\frac1{r^{2D-2}}\,W^2_\phi
\ee
and
\be
V(\phi,\chi;x)=\frac12\frac1{r^{2D-2}}\,\left(W^2_\phi+W^2_\chi\right)
\ee

The present generalization to higher dimensions is different from the
extensions one usually considers to evade the Derrick-Hobard theorem,
which include for instance constraints in the scalar fields and/or the
presence of fields with nonzero spin -- see, e.g., Ref.~{[3]}, and 
other specific works on the subject\cite{ddi,afz}. Potentials of the above
form appear for instance in the Gross-Pitaevski equation, which finds
applications in several branches of physics\cite{gp}.

\section{One Spatial Dimension}
\label{sec:one}

Here we work in $(1,1)$ space-time dimensions. We investigate models defined by
potentials that can be written as in Eqs.~(\ref{pot1})-(\ref{pot2}). Our main
motivation is to show specific features of models described by one and by two real scalar fields.

\subsection{One Field}

For models described by a single field, the equation of motion for static solutions
has the form
\be
\frac{d^2\phi}{dx^2}=W_\phi W_{\phi\phi}
\ee
We are searching for solutions that obey the boundary conditions
\be
\lim_{x\to-\infty}\phi(x)\rightarrow{\bar\phi},\;\;\;\;\;
\lim_{x\to-\infty}\frac{d\phi}{dx}\rightarrow0
\ee
where the singular point $\bar\phi$ of the potential obeys $V(\bar\phi)=0.$ 

For solutions which obey the above boundary conditions, the second-order equation of motion is equivalent to the two first-order equations \cite{bms1,bms2}
\be\label{be}
\frac{d\phi}{dx}=\pm W_\phi
\ee
These first-order equations also appear in the investigation of the energy
of static solutions. The energy density has the form
\be
\varepsilon(x)=\frac12\biggl[\left(\frac{d\phi}{dx}\right)^2+W^2_\phi\biggr]
\ee
It can be written as
\be
\varepsilon(x)=\pm\frac{dW}{dx}+\frac12\left(\frac{d\phi}{dx}\mp W_\phi\right)^2
\ee
The energy is obtained after integrating the energy density in the full real line. It is positive, and it can be minimized to the value $E=|\Delta W|=|W[\phi(\infty)]-W[\phi(-\infty)]|$ for field configurations that solve the first-order equations (\ref{be}).

Solutions of the first-order equations (\ref{be}) are named BPS and antiBPS states, and are classically or linearly stable. We show this by considering $\phi(x,t)=\phi(x)+\eta(x,t)$ and using the equation of motion to get, after discarding non-linear contributions in $\eta,$
\be
\frac{\partial^2\eta}{\partial t^2}-\frac{\partial^2\eta}{\partial x^2}+
\frac{d^2V}{d\phi^2}\biggl|_\phi\eta=0
\ee
The static field $\phi(x)$ is time-independent, then we can use
\be
\eta(x,t)=\sum_{n}\eta_n(x)\cos(w_n t)
\ee
to obtain
\be\label{se}
-\frac{d^2\eta_n}{dx^2}+U(x)\eta_n(x)=w^2_n\eta_n(x)
\ee
where $U(x)=d^2V/d\phi^2$ has to be calculated at the classical static configuration
$\phi(x)$ that describes the defect solution. The equation (\ref{se}) is a Schr\"odinger-like equation, which leads us to Quantum Mechanics. The Hamiltonian is
\be
H=-\frac{d^2}{dx^2}+U(x)
\ee
For the potential $V=(1/2)W^2_\phi,$ it can be written as
\be
H=-\frac{d^2}{dx^2}+W^2_{\phi\phi}+W_\phi W_{\phi\phi\phi}
\ee
The structure of the field-theoretical model furnishes a pair of Hamiltonians,
$H_\pm=S^\dag_\pm S_\pm,$ defined by
\be
S_\pm=\frac{d}{dx}\pm W_{\phi\phi}
\ee
which nicely lead to supersymmetric quantum mechanics\cite{khare,book}. For the issue of stability, the important point is that both $H_\pm$ are non-negative,
and so no eigenvalue in Eq.~(\ref{se}) can be negative, ensuring linear stability of the BPS states.

To illustrate the general situation, we consider the model defined by the potential
\be
V(\phi)=\frac12 \lambda^2(v^2-\phi^2)
\ee
where $\lambda$ and $v$ are parameters, real and positive. This is the
$\phi^4$ model, and here $W$ has the form
\be
W(\phi)=\lambda v^2\phi-\frac13\lambda\phi^3
\ee
with $W_\phi=\lambda(v^2-\phi^2).$ The potential has two minima, at $\bar\phi_\pm=\pm v.$ The energy of the static solutions has the form $E=(4/3)\lambda v^3.$

The model admits static solutions in the form
\be
\phi_\pm(x)=\pm v \tanh(\lambda v x)
\ee
where we are choosing $x=0$ as the center of the defect. These solutions are named kink $(+)$ and anti-kink $(-).$ They have two main features: the amplitude $(v)$ and the width $(1/\lambda v)$. Kinks and anti-kinks can be embedded in $(2,1)$ space-time dimensions, to give rise to domain ribbons, or in
$(3,1)$ dimensions, to generate domain walls. In the above example, the defect is structureless and corresponds to the Ising wall one finds in magnetic systems\cite{esche}.

In the $\phi^4$ model, the two minima defines a single topological sector. Both minima are asymmetric, that is,
they break the $Z_2$ symmetry of the model. The choice of the minimum leads to the asymmetric or ordered phase;
thus, if we introduce a mechanism to restore the symmetry, we should necessarily describe a second-order phase transition.

Another example, similar to the above one is given by the potential
\be
V(\phi)=\frac12\lambda^2(v-|\phi|)^2
\ee
where $v$ and $\lambda$ are positive. We name this the modified $\phi^2$ model. It was used in\cite{prl} as an exactly integrable model, to illustrate the presence of a single-soliton solution in a system of discrete chains.  
It can be written in terms of
\be
W=\lambda v\phi-\frac12\lambda|\phi|\phi
\ee
The first-order equations are
\be
\frac{d\phi}{dx}=\pm\lambda(v-|\phi|)
\ee
The BPS solutions are
\be
\phi(x)=\pm 2v \frac{\tanh \left( \frac{\lambda x}{2} \right)}{1+\tanh \left( \frac{\lambda |x|}{2} \right)}
\ee
There is a single topological sector, which supports BPS solutions.
The energy of the BPS sector is given by $E=\lambda v^2.$ In Fig.~[1] and [2] we plot potentials and defect solutions for both the $\phi^4$ and modified $\phi^2$ models.

\begin{figure}[ht]
\centering
\includegraphics[width=.50\textwidth]{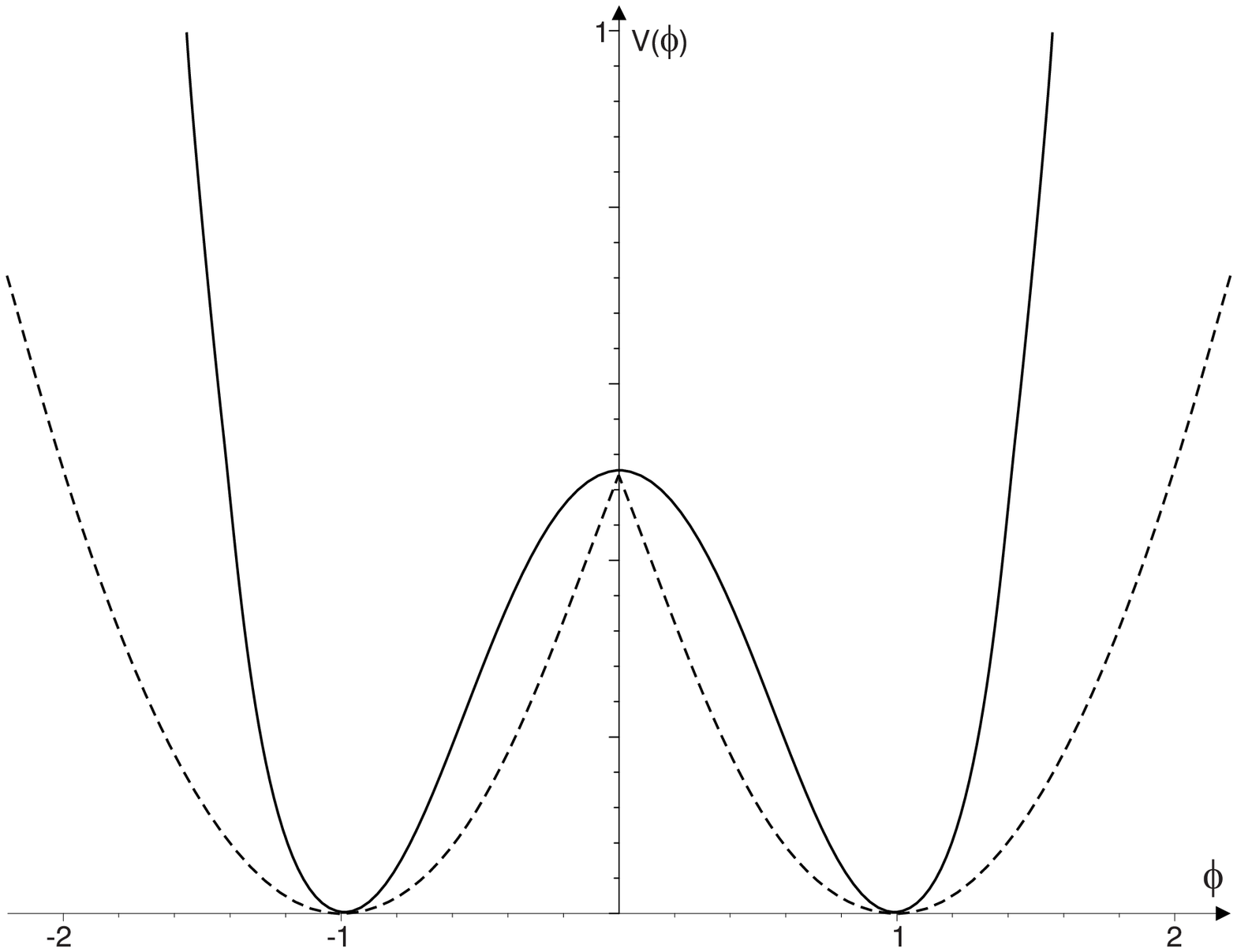}
\caption{Plots of the potentials for the $\phi^4$ and modified $\phi^2$ models,
depicted with solid and dashed lines, respectively. We use $v=1$ and $\lambda=1,$ for simplicity.}
\end{figure}

\begin{figure}[ht]
\centering
\includegraphics[width=.50\textwidth]{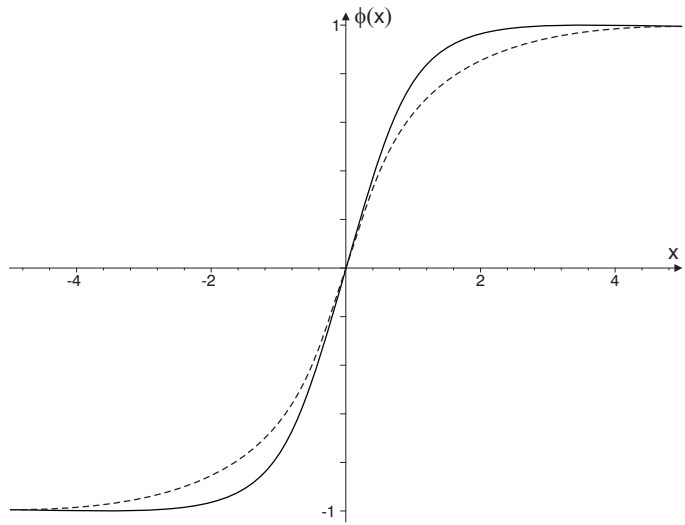}
\caption{Plots of the kink-like solutions for the $\phi^4$ and modified $\phi^2$ models, depicted with solid and dashed lines, respectively.
We use $v=1$ and $\lambda=1,$ for simplicity.}
\end{figure}

Another example is given by the potential
\be
V(\phi)=\frac12 \lambda^2\phi^2(v^2-\phi^2)^2
\ee
where $v$ and $\lambda$ are positive. This is the $\phi^6$ model, and it has three minima, one at ${\bar\phi}_0=0,$ which preserves the $Z_2$ symmetry of the model, and two at ${\bar\phi}_\pm=\pm v,$ which break the $Z_2$ symmetry. Here we have $W=\lambda v^2\phi^2/2-\lambda\phi^4/4.$ The energy of the BPS solutions is $E=(1/4)\lambda v^4.$ The solutions are
\be
\phi^\pm_\pm(x)=\pm v\sqrt{\frac12[1\pm\tanh(\lambda v x)]}
\ee
The $\phi^6$ model supports two topological sectors, degenerate in energy, defined by the minima $-1$ and $0,$ and $0$ and $1.$ Furthermore, the minimum at zero does not break the symmetry of the model, so it represents a symmetric phase. Differently from the $\phi^4$ model, the $\phi^6$ model supports two distinct but degenerate phases, one symmetric and the other asymmetric; thus, it can be used to describe a first-order phase transition.  

For this reason, it seems to be interesting to describe another model, similar to the above one, but with lower power in the field. Such a model was invented very recently in\cite{bil}, and it is governed by the potential
\be
V(\phi)=\frac12\lambda^2\phi^2(v-|\phi|)^2
\ee
where $v$ and $\lambda$ are positive. We name this the modified $\phi^4$ model, and it can be written in terms of $W=\lambda v\phi^2/2-\lambda|\phi|\phi^2/3.$ It also has three minima, one at ${\bar\phi}=0$ and two at ${\bar\phi}=\pm v.$ However, the topological solutions are different, given by
\be
\phi^\pm_\pm(x)=\pm\frac{v}{2}[1\pm\tanh(\lambda v x)]
\ee
They have energy given by $E=(1/6)\lambda v^3.$ 

In Fig.~[3] and [4] we plot the potentials and defect solutions for both the $\phi^6$ and modified $\phi^4$ models, to show how they differ from each other. There we see that the modified $\phi^4$ model is more symmetric than the former $\phi^6$ model, and may be used in applications in condensed matter as symmetrical interfaces or fronts, which can also be of interest to model periodically forced oscilatory systems governed by the Ginzburg-Landau equation\cite{ehm}.

\begin{figure}[ht]
\centering
\includegraphics[width=.50\textwidth]{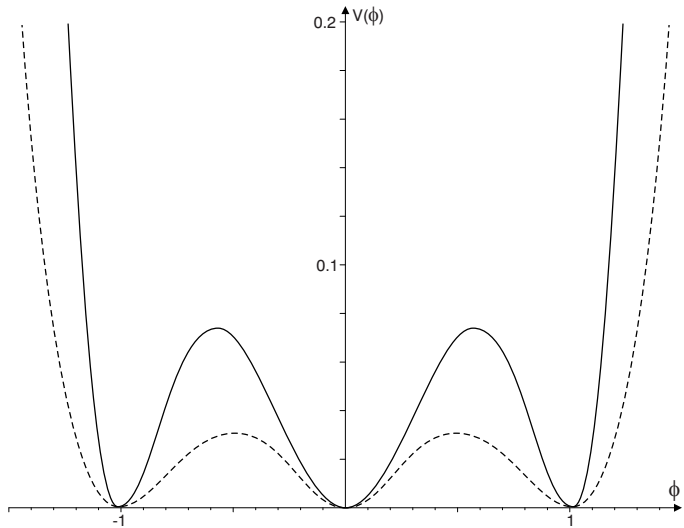}
\caption{Plots of the potentials for the $\phi^6$ and modified $\phi^4$ models,
depicted with solid and dashed lines, respectively. We use $v=1$ and $\lambda=1,$ for simplicity.}
\end{figure}

\begin{figure}[ht]
\centering
\includegraphics[width=.50\textwidth]{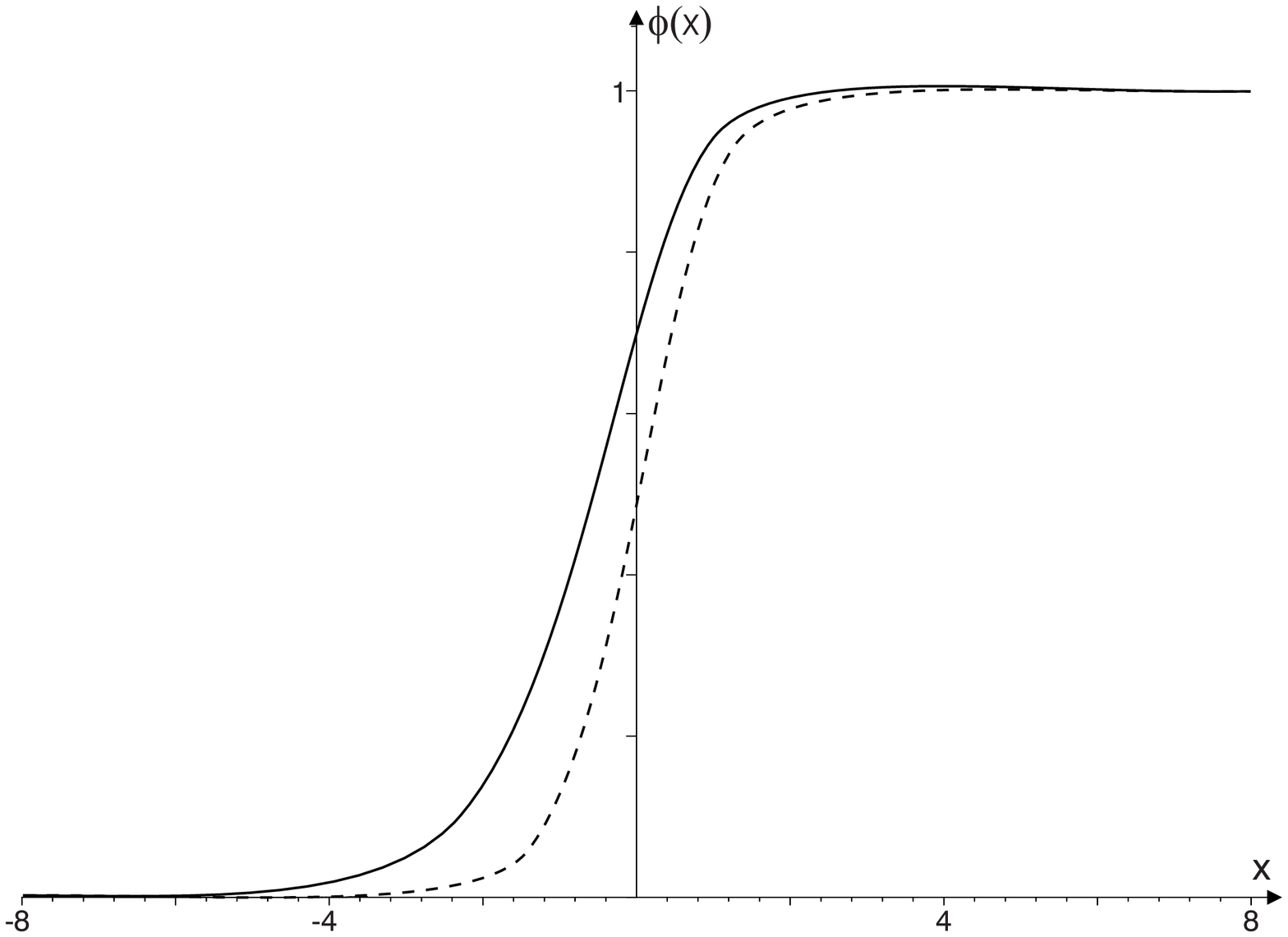}
\caption{Plots of the kinks for the $\phi^6$ and modified $\phi^4$ models,
depicted with solid and dashed lines, respectively. We use $v=1$ and $\lambda=1,$ for simplicity.}
\end{figure}

Another model is the double-sine-Gordon model. It is defined by\cite{blm}
\be
V(\phi)=\frac{2}{1+r}[r+\cos(\phi)]^2
\ee
where $r$ is a real parameter; we consider $r\in(0,1),$ and the limits $r\to0$ and $r\to1$ lead to distinct sine-Gordon models. This potential can be written in terms of
\be
W(\phi)=\frac{2}{\sqrt{1+r}}[r\phi+\sin(\phi)]
\ee

The first-order equations are given by
\be
\frac{d\phi}{dx}=\pm\frac{2}{\sqrt{1+r}}[r+\cos(\phi)]
\ee
There are two kind of solutions, representing the large and small topological sectors; there are large kinks
\be
\phi_l(x)=\pm 2 \arctan \left( \sqrt{\frac{1+r}{1-r}} \tanh \left(\sqrt{1-r} \, x\right) \right)
\ee
and small kinks
\be
\phi_s(x)=\pm \pi -2 \arctan \left( \sqrt{\frac{1-r}{1+r}} \tanh \left(\sqrt{1-r} \,x\right) \right)
\ee
Their energies are, respectively,
\be
E_l=4\sqrt{1-r}+4r\frac{\pi-\arccos(r)}{\sqrt{1+r}}
\ee
and
\be
E_s=4\sqrt{1-r}-4r\frac{\arccos(r)}{\sqrt{1+r}}
\ee

Another example is given by the family of models introduced in\cite{bmm}
\be\label{model}
V_p(\phi)=\frac12\,
\left(\phi^{\frac{p-1}{p}}-\phi^{\frac{p+1}{p}}\right)^2
\ee
Here both the field and coordinates are dimensionless. The parameter $p$ is real, and it is related to the way the field self-interacts. These models are well-defined for $p$ odd, $p=1,3,5...$ For $p=1$ we get to the standard $\phi^4$ theory. For $p=3,5,...$ we have new models, presenting potentials which support minima at $\bar\phi=0$ and $\pm1$. For $p$ odd the classical bosonic masses at the asymmetric minima $\bar\phi=\pm1$ are given by $m^2=4/p^2$. For $p=3,5,...$ another minimum appear at $\bar\phi=0$. However, the classical mass at this symmetric minimum diverges, signaling that $\bar\phi=0$ does not define a true perturbative ground state for the system.

We consider $p$ odd. The first-order equations are
\be
\frac{d\phi}{dx}=\pm\phi^{\frac{p-1}{p}}\mp\phi^{\frac{p+1}{p}}
\ee
which have solutions
\be
\phi^{(1,p)}_{\pm}(x)=\pm\tanh^p(x/p)
\ee
We consider the center of the defect at ${\bar x}=0$, for simplicity.
Their energies are given by $E^{(p)}_{1,o}=4p/(4p^2-1)$, and we plot
some of them in Fig.~[5].

\begin{figure}[ht]
\centering
\includegraphics[width=.50\textwidth]{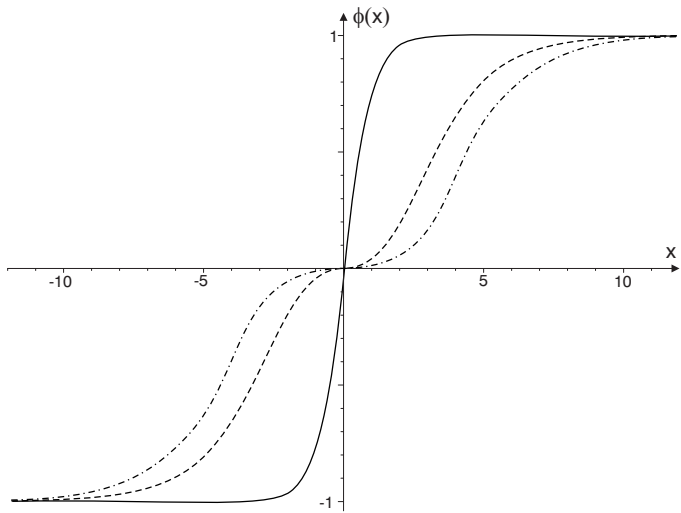}
\caption{Plots of kinks for $p=1,3,5,$ depicted with solid, dashed,
and dot-dashed lines, respectively.}
\end{figure}

We see that solutions for $p=3,5,...,$ connect the minima $\bar{\phi}=\pm1$, passing through the symmetric minimum
at $\bar{\phi}=0$ with vanishing derivative. They are new structures, which solve first-order equations, and we call them 2-kink defects since they seem to be composed of two standard kinks, symmetrically separated by a distance
which is proportional to $p$, the parameter that specifies the potential. To see this, we notice that the zero modes are given by
\be
\eta_0^{(1,p)}=c^{(1,p)}\tanh^{p-1}(x/p)\,{\rm sech}^2(x/p)
\ee
where $c^{(1,p)}=[(4p^2-1)/4p]^{1/2}$ -- we use $\eta_0^{(D,p)}$ to represent the zero modes. These zero modes concentrate around two symmetric points, which identify each one of the two standard kinks. The defect solutions
are non-perturbative, and for $p=3,5,...$ they somehow {\it know} that the minimum at zero cannot be used as a perturbative ground state.

The cases with $p$ even are different. The case $p=2$ is special: it gives
the potential 
\be
V(\phi)=\frac12\phi-\phi^2+\frac12\phi^3
\ee
which supports the non topological or lumplike solution
\be
\phi^{(1,2)}_l(x)=\tanh^2(x/2)
\ee
The lumplike solution is unstable, as we can see from the zero mode, which is proportional to $\tanh(x/2){\rm sech}^2(x/2)$: the zero mode has a node, so there must be a lower (negative energy) eigenvalue. In the present
case, the tachyonic eigenvalue can be calculated exactly since the quantum-mechanical potential associated with stability of the lumplike configuration has the form 
\be
U(x)=1-3\, {\rm sech}^2(x/2)
\ee
This potential supports three bound states, the first being a tachyonic eigenfunction with eigenvalue $w^2_0=-5/4$, the second the zero mode, $w^2_1=0,$ and the third a positive energy bound state with $w^2_2=3/4$.

The other cases for $p$ even are $p=4,6,...$ These cases require that $\phi\geq0$ in Eq.~(\ref{model}), but we can also change $\phi\to-\phi$ in Eq.~(\ref{model}) and consider $\phi\leq0$. We investigate the case $\phi\in[0,\infty)$; reflection symmetry leads to the other case. We notice that the origin is also a minimum, with null derivative.
These models also support topological defects, in the form $\phi^{(1,p)}(x)=\tanh^p(x/p)\;(x\geq0)$, with energies
$E^{(p)}_{1,e}=2p/(4p^2-1)$ for $p=4,6,...$ These solutions solve the first order equation $d\phi/dx=W_\phi$; the other equation
$d\phi/dx=-W_\phi$ is solved by $\phi(x)=-\tanh^p(x/p)\;(x\leq0)$.

In $(1,1)$ space-time dimensions, the case of $p=3,5,...$ are very interesting since they give rise to 2-kink solutions. The energy density of the 2-kink solutions are given by
\be
\varepsilon(x)=\tanh^{2p-2}(x/p)\,{\rm sech}^4(x/p)
\ee
We plot this energy density in Fig.~[6]; it is zero at the center of the defect, and it has maxima at the two symmetric points
${\bar x}_\pm=\pm p\; {\rm arcsech}\sqrt{2/(p+1)},$
showing that the 2-kink solution engenders internal structure, being composed by two kink solutions. It is very interesting to notice that this kind of behavior was recently found in a magnetic system, if one constrains the geometry of the material in a very specific way\cite{apl}.

\begin{figure}[ht]
\centering
\includegraphics[width=.50\textwidth]{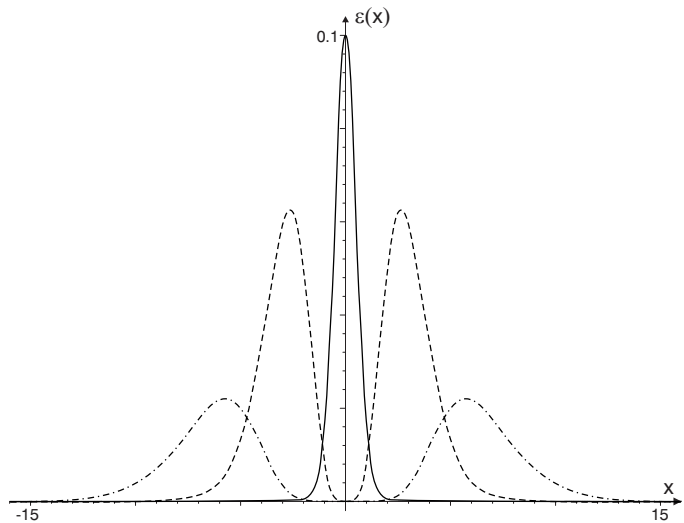}
\caption{Plots of the energy density for $p=1,3,5,$ with solid, dashed, and dot-dashed lines, respectively. The case $p=1$ is depicted with the factor $1/10$, to easy comparison.}
\end{figure}

\subsection{Two Fields}

For models described by two fields, the potential is given as in Eq.~(\ref{pot2}). The equations of motion for static solutions have the form
\ben
\frac{d^2\phi}{dx^2}&=&W_\phi W_{\phi\phi}+W_\chi W_{\chi\phi}
\\
\frac{d^2\chi}{dx^2}&=&W_\phi W_{\phi\chi}+W_\chi W_{\chi\chi}
\een
The energy density for static solutions can be written as
\ben
\varepsilon(x)&=&\frac12\left(\frac{d\phi}{dx}\right)^2
+\frac12\left(\frac{d\chi}{dx}\right)^2+\frac12W^2_\phi+\frac12W^2_\chi\nonumber
\\
&=&\pm\frac{dW}{dx}+\frac12\left(\frac{d\phi}{dx}\mp W_\phi\right)^2+
\frac12\left(\frac{d\chi}{dx}\mp W_\chi\right)^2
\een
The energy is then minimized to the value $E=|\Delta W|,$ where
\be
\Delta W=W[\phi(\infty),\chi(\infty)]-W[\phi(-\infty),\chi(-\infty)]
\ee
for field configurations which obey the first-order equations
\be\label{be2}
\frac{d\phi}{dx}=\pm W_\phi,\;\;\;\;\;\;\;\;\;\;\frac{d\chi}{dx}=\pm W_\chi
\ee

We notice that the first-order equations can be seen as dynamical systems. In this case, the singular points solve $W_\phi=0$ and $W_\chi=0,$ and so they are the absolute minima of the potential. We identify the topological sectors of the model by choosing pair of absolute minima: the pair $({\bar\phi}_i,{\bar\chi}_i)$ and $({\bar\phi}_j,{\bar\chi}_j)$ identify the sector $(ij),$ with energy $E^{(ij)}.$ If
$W({\bar\phi}_i,{\bar\chi}_i)\neq W({\bar\phi}_j,{\bar\chi}_j),$ the sector $(ij)$
is a BPS sector, with energy given by
$E^{(ij)}=|W({\bar\phi}_i,{\bar\chi}_i)-W({\bar\phi}_j,{\bar\chi}_j)|.$ In this case,
the first-order equations (\ref{be2}) have non-trivial solutions which connect the minima $({\bar\phi}_i,{\bar\chi}_i)$ and $({\bar\phi}_j,{\bar\chi}_j).$

To exemplify the general situation, let us consider the model defined by
\be
W=\lambda v^2\phi-\frac13\lambda\phi^3-\mu\phi\chi^2
\ee
This model was first investigated in\cite{bds}, and it can be seen as an extension of the $\phi^4$ model considered in the former subsection. The first-order equations are
\be
\frac{d\phi}{dx}=\pm\lambda(v^2-\phi^2)\mp\mu\chi^2,\;\;\;\;\;
\frac{d\chi}{dx}=\mp2\mu\phi\chi
\ee
For $\lambda/\mu>0,$ the absolute minima are $v_{1,2}=(\pm v,0)$ and
$v_{3,4}=(0,\pm v\sqrt{\lambda/\mu}).$  There are five BPS sectors, identified
by the energies: $E^{(12)}=(4/3)|\lambda v^3|$ and $E^{(13)}=E^{(14)}=E^{(23)}=E^{(24)}=(2/3)|\lambda v^3|.$ Also, there is a single non-BPS sector, identified by the minima $v_3$ and $v_4.$

In the sector defined by the minima $v_{1,2}$ we have several BPS solutions. Some of them are given by
\be
\phi_\pm(x)=\pm v\tanh(\lambda v x),\;\;\;\;\;\chi(x)=0
\ee
They describe a straight line segment connecting the two minima $(\pm v,0)$ in the $(\phi,\chi)$ plane. Other pairs of solutions are
\be
\phi_\pm(x)=\pm v\tanh(2\mu vx),\;\;\;\;\;
\chi(x)=\pm v\sqrt{\lambda/\mu-2}\;{\rm sech}(2\mu vx)
\ee
for $\lambda/\mu\geq2.$ They describe eliptycal arcs connecting the minima $v_{1,2}=(\pm v,0).$ These solutions
describe orbits in the configuration space which obey 
\be
\phi^2+\frac{1}{\lambda/\mu-2}\chi^2=v^2
\ee
We notice that in configuration space, the one field solutions describe straight line orbits, and the two field solutions describe eliptycal orbits. They very much remind us of linearly and eliptycally polarized light.
This scenario is interesting, and may be used to model more sophisticated problems. A good example can be found in works used to describe chiral interface\cite{chiral1,chiral2}. 

The presence of the second field $\chi$ leads to the interesting situation: For $x=0,$ in the center of the defect
the second field does not vanishes, and may be used to describe internal structure. The profile of the wall the model generates is the same of finds in the so called Bloch wall scenario, which appear in models described by the Ginzburg-Landau equation, as one finds in magnetic systems\cite{esche} and in other situations\cite{wa}.   

Models described by two or more fields are also used to describe conformational structures in almost one dimensional polimeric chains. Work in this line has been done in several different contexts,
in particular to describe defect structure in a ferroelectric crystal\cite{pma,brs} and in polyethylene\cite{bve,vsb}.
Other form of interactions can be of use to describe conductivity in Langmuir films\cite{bll}.

The presence of internal structure can be extended to describe scenarios
where a single field entraps two fields, generating more sophisticated structures, where networks of domain walls live inside a given domain wall. This idea was first used to simulate nanotubes, fullerenes
and soliton stars\cite{bbrc,bb01,sut}. 

\section{Two or More Spatial Dimensions}
\label{sec:two}

To find defect structures in higher dimensions, we consider the class of potentials given by Eq.~(\ref{potD}). In this case, we suppose the model supports static and radial solutions, in the form $\phi=\phi(r).$
This leads to the equation of motion
\be
\frac1{r^{D-1}}\frac{d}{dr}\left(r^{D-1}\frac{d\phi}{dr}\right)=\frac1{r^{2D-2}}\,W_\phi W_{\phi\phi}
\ee
which can be written as
\be\label{em}
r^{D-1}\frac{d}{dr}\left(r^{D-1}\frac{d\phi}{dr}\right)=W_\phi W_{\phi\phi}
\ee
We change $r\to y,$ obeying $r^{D-1}(d/dr)=(d/dy)$ to get
\be
\frac{d^2\phi}{dy^2}=W_\phi W_{\phi\phi}
\ee
which is the equation of motion for static solutions in $D=1.$ Thus, we reduce the problem of finding static solutions of some $(D,1)$ dimensional model, to that of effectively finding static solutions of some much simpler $(1,1)$ dimensional model.

The energy of the static, spherically symmetric field configuration has the form
\be
E=\frac12\Omega_D \int_0^\infty r^{D-1} dr
\biggl[\left(\frac{d\phi}{dr}\right)^2+\frac1{r^{2D-2}}W^2_\phi\biggr]
\ee
where $\Omega_D$ is the angular factor: $\Omega_D=2\pi^{D/2}\Gamma(D/2),$ with $\Gamma$ standing for the gamma function. We can work with the energy to write
\be
E=\frac12\Omega_D\int_0^\infty r^{D-1}dr\left(\frac{d\phi}{dr}\mp\frac1{r^{D-1}W_\phi}\right)^2
\pm\Omega_D\int_0^\infty dr\;\frac{dW}{dr}
\ee
It can be minimized to the value $E_{BPS}=\Omega_D|\Delta W|,$ where $\Delta W=W(\phi(r\to\infty))-W(\phi(r\to0)),$
for field configurations that obey
\be\label{beD}
\frac{d\phi}{dr}=\pm\frac1{r^{D-1}} W_\phi
\ee

The above results are extensions to $D\geq2$ of results valid in $(1,1)$ space-time dimensions. To get to them, an important manipulation was the change of variables $r\to y;$ for $D=2$ it leads to $y=\pm\ln(r),$ which relates the interval $[0,\infty)$ to $(-\infty,\infty).$ However, for $D\geq3,$ one gets $y=\pm r^{2-D}/(2-D),$ and now $y$ must be in
the interval $[0,\infty)$ or $(-\infty,0].$ Thus, for $D\geq3,$ the unidimensional problem that one has to solve is different from the standard situation, since the coordinate $y$ does not span the full real line anymore. For this reason, to work in higher $(D\geq3)$ dimensions, one has to invent new models, which need to be solved in the interval $[0,\infty)$.

The presence of first-order equations corresponds to the Bogomol'nyi bound, now generalized to models for scalar field
that live in $D$ spatial dimensions. The result shows that the solutions obey first order equations and have energy evenly distributed into gradient and potential portions. We notice that for $D=2$ the model engenders scale invariance. Evidently, the solutions of the above first order equations solve the equation of motion (\ref{em}) for potentials given by Eq.~(\ref{potD}). Also, we follow former work\cite{bms1} and introduce the ratio $R=r^{D-1}(d\phi/dr)/W_\phi$. For field configuration that obey $\phi(0)=\bar{\phi}$ and $\lim_{r\to0}d\phi/dr\to0$ one can show that solutions of the equation of motion also solve the first-order equations (\ref{beD}). This extends former result\cite{bms1} to the present investigation: it shows that the equation of motion (\ref{em}) completely factorizes into the two first
order equations (\ref{beD}). Thus, all the topological solutions are BPS states.

The equation of motion for $\phi=\phi(r)$ is
\be\label{em1}
\frac1{r^{D-1}}\frac{d}{dr}
\left(r^{D-1}\frac{d\phi}{dr}\right)=\frac1{r^{2D-2}}W_\phi W_{\phi\phi}
\ee
The first order equation is given by (\ref{beD}), and their solutions solve the equation of motion and are stable against radial, time-dependent fluctuations. To see this explicitly, we consider $\phi(r,t)=\phi(r)+\sum_k \eta_k(r) \cos(w_{k}\, t)$. For small fluctuations we get $H\eta_k=w^2_{k}\eta_k$, where the Hamiltonian can be written as
\be
H=\frac{1}{r^{2D-2}}\left(-r^{D-1}\frac{d}{dr}\mp W_{\phi\phi}\right)
\left(r^{D-1}\frac{d}{dr}\mp W_{\phi\phi}\right)
\ee
It is non-negative, and the lowest bound state is the zero mode, which obeys $r^{D-1}d\eta_0/dr=\pm\,W_{\phi\phi}\,\eta_0.$ This gives
$\eta_0(r)=c\,\exp\left({\pm\int dr \,{r^{1-D}}\,W_{\phi\phi}}\right),$
where $c$ is the normalization constant, which usually exists only for one of the two sign possibilities. We can also write $\eta_0(r)=c W_\phi$, which is another way to write the zero mode.

In $D=1$ one usually introduces the conserved current $j^\mu=\varepsilon^{\mu\nu}\partial_\nu\phi$. We see that $\rho=d\phi/dx$ and so $\rho^2$ gives the energy density of the field configuration\cite{bb}.
Thus, we introduce 
\be\label{tc}
Q_T=\int^{\infty}_{-\infty}dx \rho^2
\ee
as  the topological charge, which is exactly the total energy of the solution.

We now consider $D=2$. The equation of motion is $\nabla^2\phi=(1/r^2)W_\phi W_{\phi\phi}$. We search for solution $\phi(r)$, which only depends on the radial coordinate, obeying $\phi(0)=\bar{\phi}$ and $\lim_{r\to0}d\phi/dr\to0$.
In this case we get
\be
r\frac{d}{dr}\left(r\frac{d\phi}{dr}\right)=
\frac{dW}{d\phi}\frac{d^2W}{d\phi^2}
\ee 
We write $dx=\pm r^{-1}dr$ to get $d^2\phi/dx^2=W_\phi W_{\phi\phi}.$
This result maps the $D=2$ model into the $D=1$ model.

We see that $r=\exp(\pm x),$ which shows that the full line
$x\in(-\infty,\infty)$ is mapped to $r\in[0,\infty)$. We notice
that since the center of the defect is arbitrary, so it is the point $r=1$
in $D=2$; thus the solution introduces no fundamental scale,
in accordance with the scale symmetry that the model engenders. If one uses
the model (\ref{model}) with $p$ odd to define the potential in this case,
we get the solutions
\be
\phi_{\pm}^{(2,p)}(r)=
\pm\left(\frac{r^{2/p}-1}{r^{2/p}+1}\right)^p
\ee
Their energies are $E^{(p)}_2={8\pi p}/(4p^2-1)$. In Fig.~[7] we depict
the defect solution for $p=1$. The corresponding zero mode is given by
$\eta^{(2,1)}_0(r)=\sqrt{32/\pi}[r^2/(r^2+1)^2].$
This zero mode binds around the circle where the defect solution vanishes,
as we show in Fig.~[7], where we also plot
$\rho^{(2,1)}_0(r)=32r^4/\pi(r^2+1)^4$.

\begin{figure}[ht]
\centering
\includegraphics[width=.60\textwidth]{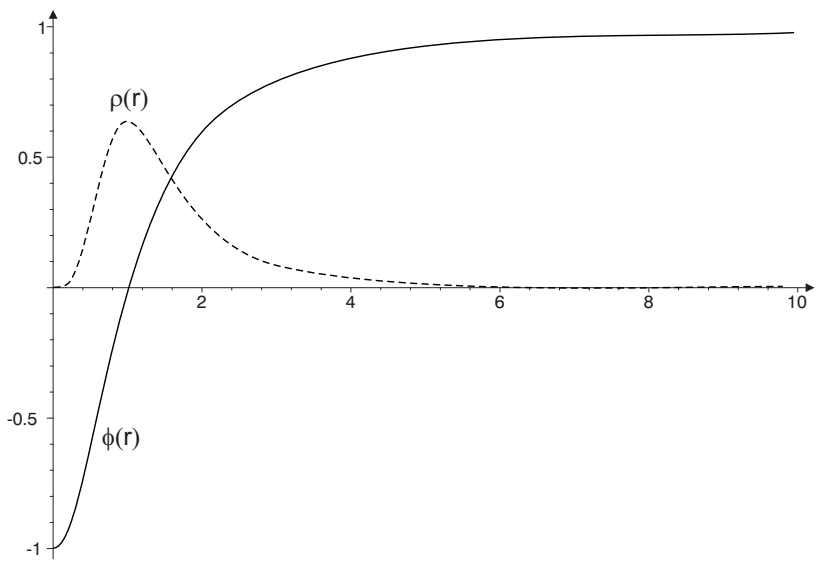}
\caption{Radial defect solution $\phi(r)$ and density $\rho(r)$ of the
corresponding zero mode for $D=2$, in the case $p=1.$}
\end{figure}

We now investigate a model described by a complex scalar field. Here the equation of motion has the form
\be
\nabla^2\varphi=\frac1{r^2}\frac{\partial V}{\partial{\bar\varphi}}
\ee
where bar indicates complex conjugation. We suppose that the field configuration has the form $\varphi(r,\theta)=\phi(r)e^{im\theta},$ for $m=1,2,3,...$ In this case we get
\be
r\frac{d}{dr}\left(r\frac{d\phi}{dr}\right)=\frac{dV_{eff}}{d\phi}
\ee
where
\be\label{peff}
V_{eff}(\phi)=V(\phi)+\frac12\,m^2\phi^2
\ee

In view of the effective potential in (\ref{peff}), if one starts with
\be
V(\phi)=\frac12\phi^6-A\phi^4
\ee
one gets to
\be\label{potd26}
V_{eff}(\phi)=\frac12\phi^6-A\phi^4+\frac12m^2\phi^2
\ee
This is the $\phi^6$ potential -- see Eq.~(27); thus, it has the property that, if $A$ is integer, the value $m_c=A$ shows that there will always be a critical $m_c$ for which there are analytical solutions representing topological solitons in the model. Such solutions are given by
\be
\phi^{\pm}_{\pm}(r)=\pm\frac{A}{\sqrt{1+r^{\pm2A}}}
\ee
Moreover, for $m<m_c$ one can find non-topological solutions numerically; in Fig.~[8] we show the potential for $A=2$ and for $m=0,1,2$. The inset shows how the potentials behave for $\phi$ small: there are no lumplike solutions for $m=0$ or $2,$ but for $m=1.$

\begin{figure}[ht]
\centering
\includegraphics[width=.60\textwidth]{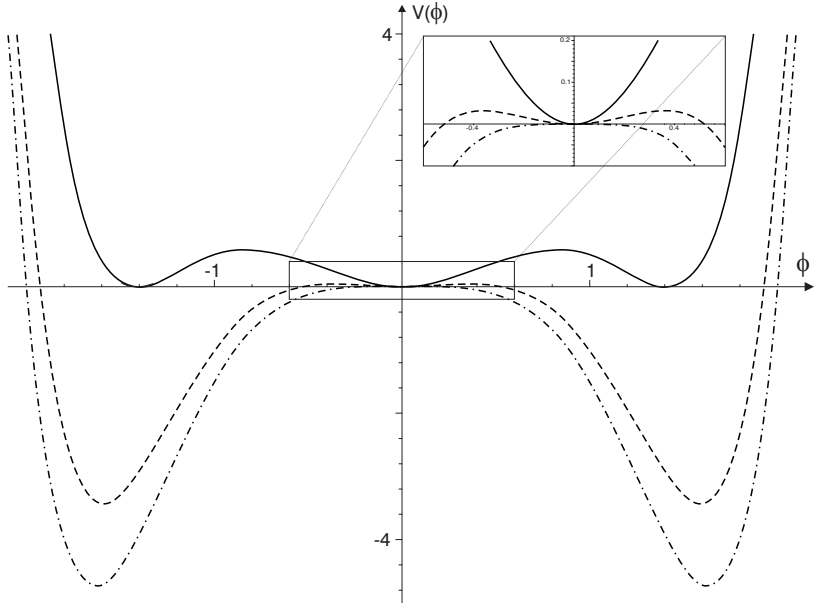}
\caption{Plots of the potential in Eq.~(\ref{potd26}) for $A=2,$ and for $m=2,1,0,$ depicted as solid, dashed and dot-dashed lines, respectively. The inset details
the behavior for $\phi$ small.}
\end{figure}

We now return to models described by a single real scalar field, and we consider the case $D\geq3$. The equation of motion for $\phi=\phi(r)$ is
\be
r^{D-1} \frac{d}{dr}\left(r^{D-1}
\frac{d\phi}{dr}\right)=W_\phi W_{\phi\phi}
\ee
We write $dx=\pm r^{1-D}dr$ to get $d^2\phi/dx^2=W_\phi W_{\phi\phi}$ and again, we map the model into a one-dimensional problem. We solve $dx=\pm r^{1-D}dr$ to get $x=\mp r^{(2-D)}/(D-2)$. This shows that
$x$ is now in $(-\infty,0]$ or $[0,\infty)$, and so we have to use the upper sign for $x\leq0$, or the lower sign for $x\geq0$. We can use the model of Eq.~(\ref{model}) for $p=4,6,...$ to solve the D-dimensional problem with $N=2D-2$. In this case the solutions are
\be
\phi^{(D,p)}(r)=\tanh^{\,p}\biggr[\,\frac1p\,
\left(\frac{r^{2-D}}{D-2}\right)\biggl]
\ee
for $D=3,4,...$ Their energies are given by $E_D^{(p)}=\Omega_D[{2p}/(4p^2-1)].$

In the case of $D=3$ and $p=4$ the zero mode is given by $\eta_0^{(3,4)}(r)=c^{(3,4)}\tanh^3(1/4r){\rm sech}^2(1/4r);$
we could not find the normalization factor explicitly in this case. The zero mode binds at the skin of the defect solution, concentrating around the radius $R$ of the defect, which is given by $R=1/4\,{\rm arctanh}[(1/2)^{1/4}]$.

We further illustrate the case $D=2$ with another very interesting model, described by the potential
\be\label{sg2}
V(\phi)=\frac12\frac{1}{r^2}\sin^2(\phi)
\ee
The equation of motion can be written as
\be
r\frac{d}{dr}\left(r\frac{d\phi}{dr}\right)=\sin(\phi)\cos(\phi)
\ee
or, using $r(d/dr)=d/dy,$
\be
\frac{d^2\phi}{dy^2}=\sin(\phi)\cos(\phi)
\ee
The first-order equations are
\be
\frac{d\phi}{dy}=\pm\sin(\phi)
\ee
This is the sine-Gordon model, and the soliton solution has the form
$\phi(y)=\pm2\arctan(e^{\pm y}).$ If we turn back to the radial coordinate, we get
$\phi(r)=\pm2\arctan(r^{\pm1}),$ which is depicted in Fig.~[9]. The model maps an old solution, shown in Ref.~{\cite{bp}}. It reproduces topological solitons in magnetic systems, as one can see in
Ref.~{\cite{rus}}. The same solution appears in the more recent work\cite{bog02}, in the form of delocalized vortex structure in the spin-flop
phase in crystals with the $C_{nv}$ symmetry. Our solution shown in Fig.~[9] maps the angle formed between the staggered vector ${\vec l}$ and the tetragonal axis, as it is shown in Fig.~[12] of Ref.~{\cite{bog02}}.

\begin{figure}[ht]
\centering
\includegraphics[width=.60\textwidth]{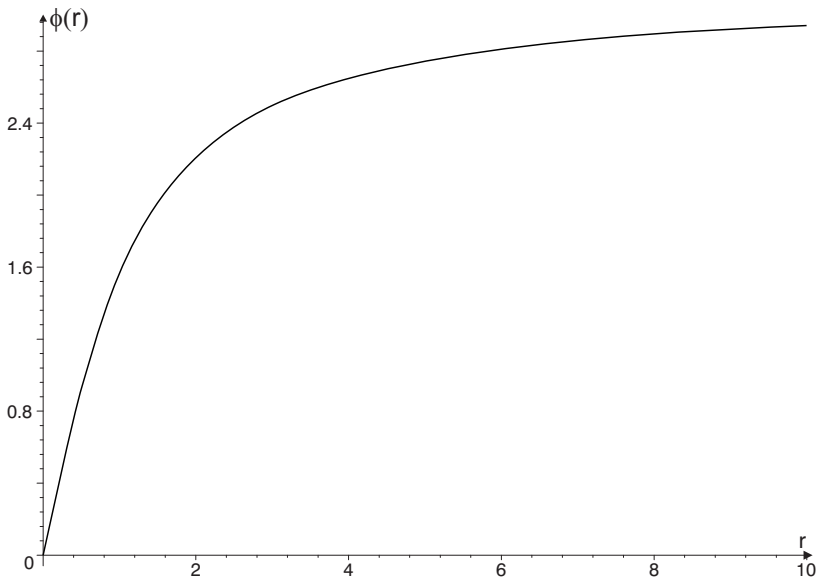}
\caption{Plot of the radial solution $\phi(r)=2\arctan(r),$ which represents a topological soliton for the model described by the potential of Eq.~(\ref{sg2}).}
\end{figure}

The solutions that we have found for $D\geq2$ have a central core, and a skin which depends on the parameters that specify the potential of the model. They are stable, well distinct of other known global defects such as the bubbles formed from unstable domain walls\cite{b1,b2,b3,b4}. Also, they are neutral structures, and may contribute to curve space time, and to affect cosmic evolution. They may become charged if charged bosons and/or fermions bind to them. For instance, if one couples Dirac fermions with the Yukawa coupling $Y(\phi)=r^{1-D}\,W_{\phi\phi}$, the fermionic zero modes are similar to the bosonic zero modes that we have just presented.

We notice that we have solved the equation of motion using
$r^{1-D}dr=\pm dx$. Although this identification works very naturally for
the first order equations, we used the equation of motion to present a more
general investigation, which may be extended to tachyons in $D$ spatial
dimensions. For instance, from the lumplike solution for $D=1$ and $p=2$
we can obtain another lumplike solution $\phi^{(2,2)}_l(r)=(r-1)^2/(r+1)^2,$
which is valid for $D=2$. Another issue concerns the identification of the
topological behavior of the above defect structures. We do this introducing
the (generalized current-like) tensor
\be
j^{\mu_1\mu_2\cdots\mu_{D}}=
\varepsilon^{\mu_1\mu_2\cdots\mu_{D}\mu_{D+1}}\partial_{\mu_{D+1}}\phi
\ee
It obeys $\partial_{\mu_i}j^{\mu_1\mu_2\cdots\mu_{D}}=0$
for $i=1,2,\cdots,D,$ which means that the quantities
\be
\rho^{i_1i_2\cdots i_{D-1}}=j^{0i_1i_2\cdots i_{D-1}}
\ee
constitute a family of $D$ distinct conserved (generalized charge)
densities. We introduce the scalar quantity 
\be
\rho_D^2=\rho_{i_1i_2\cdots i_{D-1}}\rho^{i_1i_2\cdots i_{D-1}}=(-1)^D(D-1)!
(d\phi/dr)^2
\ee
which generalizes the standard result, obtained with
$j^\mu=\varepsilon^{\mu\nu}\partial_\nu\phi$ -- see the reasoning
leading to Eq.~(\ref{tc}). Thus, we define the
topological charge as
\be
Q^D_T=\int d{\vec r}\,\rho_D^2=(-1)^D(D-1)!\Omega_D\Delta W
\ee
which exposes the topological behavior of the new global defect
structures that we have just found.

The authors would like to thank financial support from PROCAD/CAPES and PRONEX/CNPq/FAPESQ. DB thanks CNPq for partial support, and JM and RM thank CAPES for fellowships.


\end{document}